\documentclass[aps,reprint,superscriptaddress]{revtex4-2}

\usepackage{graphicx}
\usepackage{dcolumn}
\usepackage{bm}
\usepackage{xcolor}
\usepackage{float}
\usepackage{verbatim}
\usepackage{soul}

\usepackage{amsmath,amssymb}

\begin{document}


\title{Telecom quantum memory over one microsecond in nanophotonic lithium niobate}

\author{Priyash Barya}
\thanks{These authors contributed equally.}
\affiliation{Department of Electrical and Computer Engineering, University of Illinois Urbana-Champaign, Urbana, IL 61801 USA}
\affiliation{Holonyak Micro and Nanotechnology Laboratory, University of Illinois Urbana-Champaign, Urbana, IL 61801 USA}

\author{Daren Chen}
\thanks{These authors contributed equally.}
\affiliation{Holonyak Micro and Nanotechnology Laboratory, University of Illinois Urbana-Champaign, Urbana, IL 61801 USA}
\affiliation{Department of Physics, University of Illinois Urbana-Champaign, Urbana, IL 61801 USA}

\author{Ashwith Prabhu}
\affiliation{Holonyak Micro and Nanotechnology Laboratory, University of Illinois Urbana-Champaign, Urbana, IL 61801 USA}
\affiliation{Department of Physics, University of Illinois Urbana-Champaign, Urbana, IL 61801 USA}

\author{Laura Heller}
\affiliation{Department of Electrical and Computer Engineering, University of Illinois Urbana-Champaign, Urbana, IL 61801 USA}
\affiliation{Holonyak Micro and Nanotechnology Laboratory, University of Illinois Urbana-Champaign, Urbana, IL 61801 USA}

\author{Edmond Chow}
\affiliation{Holonyak Micro and Nanotechnology Laboratory, University of Illinois Urbana-Champaign, Urbana, IL 61801 USA}

\author{Hansol Kim}
\affiliation{Department of Physics, University of Illinois Urbana-Champaign, Urbana, IL 61801 USA}

\author{Joshua Akin}
\affiliation{Department of Electrical and Computer Engineering, University of Illinois Urbana-Champaign, Urbana, IL 61801 USA}
\affiliation{Holonyak Micro and Nanotechnology Laboratory, University of Illinois Urbana-Champaign, Urbana, IL 61801 USA}

\author{Vasileios Niaouris}
\affiliation{Q-NEXT, Argonne National Laboratory, Lemont, IL 60439 USA}

\author{Jiefei Zhang}
\affiliation{Q-NEXT, Argonne National Laboratory, Lemont, IL 60439 USA}
\affiliation{Materials Science Division, Argonne National Laboratory, Lemont, IL 60439 USA}

\author{Alan M. Dibos}
\affiliation{Q-NEXT, Argonne National Laboratory, Lemont, IL 60439 USA}
\affiliation{Center for Molecular Engineering, Argonne National Laboratory, Lemont, IL 60439 USA}

\author{Pengjie Wang}
\affiliation{Department of Physics, University of Illinois Urbana-Champaign, Urbana, IL 61801 USA}

\author{Elizabeth A. Goldschmidt}
\altaffiliation{goldschm@illinois.edu}
\affiliation{Department of Electrical and Computer Engineering, University of Illinois Urbana-Champaign, Urbana, IL 61801 USA}
\affiliation{Holonyak Micro and Nanotechnology Laboratory, University of Illinois Urbana-Champaign, Urbana, IL 61801 USA}
\affiliation{Department of Physics, University of Illinois Urbana-Champaign, Urbana, IL 61801 USA}

\begin{abstract}
Nanophotonic quantum memory is a vital component for scalable quantum information processing for quantum computing, networking, and sensing applications. We store single-photon-level telecom-band optical pulses for more than 1~\textit{\textmu}s using an atomic frequency comb in erbium-doped thin-film lithium niobate, well beyond what is practically feasible via propagation in even the best nanophotonic devices due to propagation losses. We verify the quantum nature of this storage by demonstrating the phase coherence and sub-single-photon noise upon retrieval. We also show the flexibility of our platform by storing up to 20 temporal modes and demonstrating an acceptance bandwidth up to 2.2~GHz. These results establish erbium-doped thin-film lithium niobate as a practical platform for on-chip quantum memory at telecom wavelengths, a key missing element for photonic quantum computing and quantum networking.
\end{abstract}

\maketitle


Quantum information systems including quantum computers \cite{Arute2019, Bluvstein2023, Kim2023}, quantum networks \cite{Kimble2008, Wehner2018}, and quantum sensors  \cite{Degen2017} have the potential to revolutionize how we process and share information. Many such systems rely on photons encoded as qubits to carry quantum information from one place to another \cite{Flamini2019}, whether just across a single chip \cite{OBrien2009} or around the world through an optical fiber or satellite link \cite{Liao2017, Yin2017}. The inability to store photons for extended periods of time presents a major barrier to scalable quantum networking and modular or distributed quantum computing or sensing. Propagation losses currently limit storage of photons to a few tens of \textit{\textmu}s in free space \cite{arnold2024all} or optical fiber and a few ns in nanophotonic devices \cite{zhang2019electronically}. Overcoming these limits requires a quantum memory that transfers the photon into the state of one or many coherent emitters in a reversible way that enables retrieval at a later time \cite{lvovsky2009optical, afzelius2009multimode}. Storage time in such quantum memories can far exceed the limits on photon storage imposed by propagation losses. Rare-earth emitters embedded in solid-state hosts have emerged as leading candidates for quantum memory due to their exceptionally long optical and spin coherence times, their compatibility with integrated photonic platforms \cite{zhong2017nanophotonic, dutta2019integrated} and, in the case of erbium, a transition in the telecommunications band \cite{wang2022er, thiel2010optical, barya2025ultra}. The atomic frequency comb (AFC) protocol for quantum memory is particularly well-suited for ensembles of rare-earth emitters and has been demonstrated in several different bulk \cite{de2008solid, lauritzen2010telecommunication, pearson2025narrow, zhang2023telecom} and integrated \cite{zhong2017nanophotonic, dutta2023atomic} platforms. Efficient quantum storage of single photon-level signals in a nanophotonic platform with the potential for wafer-scale integration, however, remains an outstanding challenge \cite{rinner2026quantum}.

At the same time, thin-film lithium niobate (TFLN) has emerged as a leading platform for quantum nanophotonics due to its wide bandgap and large $\chi^{(2)}$ nonlinearity, which is suitable for generating, controlling, and transducing quantum states of light \cite{hu2025integrated, karami2026integrated}. In addition, TFLN supports the fabrication of low-loss nanophotonic devices [13, 14] and offers strong potential for wafer-scale processing \cite{luke2020wafer}. It has also been shown to host rare-earth emitters with promising coherence times and optical properties \cite{barya2025ultra, dutta2019integrated, wang2022er}. Recent progress demonstrating narrowband spectral filtering \cite{zhao2024cavity}, cavity narrowing \cite{barya2025ultra}, AFC storage in the classical regime \cite{dutta2023atomic}, and single emitter addressing \cite{yang2023controlling} have demonstrated the potential of this platform for quantum photonic applications.

\begin{figure*}[t]
\includegraphics[width=\linewidth,keepaspectratio]{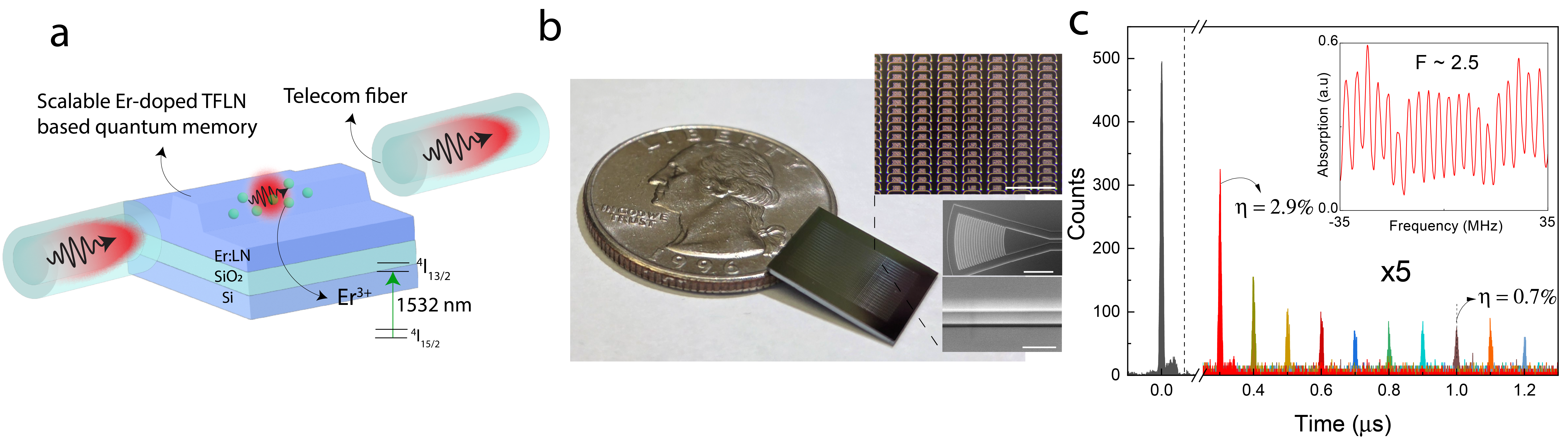}
\caption{\label{fig:QM}  \textbf{Quantum Memory in TFLN.} (a) Schematic of our Er-doped TFLN waveguides for quantum memory. (b)  Our chip-scale device. The top inset (scale bar - 500 $\mu$m) shows numerous repeating functional waveguides, highlighting its scalability. The middle (scale bar - 5 $\mu$m) and bottom (scale bar - 1 $\mu$m) insets are scanning electron microscope images of a fabricated grating coupler and waveguide respectively. (c) AFC storage of weak coherent pulses ($\bar{n}=1$) for ten different storage times from 300~ns to 1.2~$\mu\rm{s}$ spaced at 100~ns (different colors). The undelayed leakage light is shown for reference at time zero (with 5x higher scale). Inset: Atomic frequency comb for 300~ns storage time with finesse $F\approx2.5$  measured via transmission of weak probe light. }
\end{figure*}

Here, we realize an AFC memory in an erbium-doped TFLN nanophotonic waveguide to store
telecom-band photons for more than 1 $\mu$s. We observe retrieval efficiencies as high as 2.9\% for a 300~ns delay, broadband storage of pulses as wide as 2.2~GHz, and storage of up to 20 temporal modes. We verify the quantum nature of the storage by measuring a lower bound on our retrieval fidelity of 96.6\% for stored time-bin qubits and added noise equivalent that is less than $0.2$~photons (noise added per retrieved photon \cite{heller2022raman}) for all delays investigated. Achieving an equivalent delay by propagation would require a $\approx150~\rm{m}$-long waveguide which, in addition to being impractical to fabricate, would limit transmission efficiency to less than $10^{-15}$ under the most optimistic assumptions of 1 dB/m of propagation loss \cite{liu2021high, zhu2024twenty}. Storing photons using AFC in TFLN devices could open up new possibilities for synchronization and multiplexing of on-chip photon pair sources \cite{luo2017chip}, feedback and feed-forward operations in photonic quantum information processing \cite{pichler2017universal, bartlett2021deterministic}, entanglement distribution across optical fiber \cite{hensen2015loophole}, and more.

\section*{Results}
\subsection*{Quantum memory in TFLN}
Implementing AFC storage involves preparing a comb structure in the absorption profile of the erbium ensemble via spectral hole burning to remove population in a periodic way. The spacing of the comb teeth in frequency, $\Delta$, sets the storage time (or delay time) $\tau=\Delta^{-1}$, of an incident photon \cite{afzelius2009multimode}. Rare-earth ensembles are particularly suited to this because their combination of narrow optical homogeneous linewidths, large optical inhomogeneous linewidths, and long-lived spin states enable high-resolution spectral hole burning to create AFC structures by simply sending a train of pulses separated by $\tau$. There are several metrics to consider when evaluating the usefulness of quantum memory, primarily storage time, storage efficiency, acceptance bandwidth, and added noise. AFC storage is appealing because it enables competitive or leading values in each of these metrics. One challenge in any quantum memory demonstration is finding and implementing a regime that balances these, as there are many tradeoffs between different metrics. For instance, the doping concentration affects the homoegeneous linewidth, the spectral hole lifetime, and the optical density, which together affect the storage time, retrieval efficiency, and noise floor. We choose a high doping concentration (0.5\%) since it provides high optical density, which is the most direct route to efficient storage. The higher doping increases neighboring Er interactions, which we mitigate by using optimal sample conditions, as described below.

We prepare an AFC in a 0.6~mm-long ridge waveguide fabricated on 625~nm thick TFLN (see fig. \ref{fig:QM}a-b and Methods for details), where optical confinement is achieved by the lower refractive-index contrast provided by a SiO$_2$ layer below the waveguide and air above. Light is delivered to the waveguide from optical fiber into grating couplers and the device is loaded into a dilution refrigerator operating at a base temperature of 30~mK, with a magnetic field of 1.5~T applied along the $c$-axis of LN (Supplementary Figure S1). This field is essential for enabling persistent spectral holes by freezing out neighboring spin flips and extending the lifetime of the ground state levels \cite{thiel2010optical}. This ensures a low noise floor because it enables us to burn the comb and then wait several excited state lifetimes before sending in the pulse for storage, thus avoiding the spontaneous emission background that would be present if the holes were much shorter lived. To couple light in and out of the device, we developed a fiber-gluing procedure that reliably survives cooldown to cryogenic temperatures (Supplementary Figure S1). We measured the optical coherence time of the Er ensemble under these conditions to be $T_2 = 90~\mu$s, giving a homogeneous linewidth of $\gamma_h=1/\pi T_2=3.5~\rm{kHz}$ (Figure S2).

To prepare the AFC, we send a train of gaussian pulses each with a full-width-half maximum (FWHM) of 8.7~ns  (Figure S1) tuned near the 1532-nm absorption peak of erbium. The pulse separation in time, $\tau$, determines the AFC storage time, allowing us to program different delays. We verified comb preparation by measuring transmission of a weak probe through the waveguide (sent 15 ms after the burn sequence) as we sweep its frequency across the comb. For a programmed storage time of $\tau=300~\rm{ns}$, we observe a comb with $\Delta=\tau^{-1}=3.3~\rm{MHz}$ tooth spacing with a finesse of 2.5 over a 50~MHz bandwidth (Figure \ref{fig:QM}c inset). The comb finesse, defined as $F=\Delta/\gamma$ where $\gamma$ is the width of each comb tooth, impacts the storage efficiency, $\eta$, which should approach the theoretical value \cite{de2008solid, afzelius2009multimode}:
\begin{align}
\eta= \left(\frac{\text{OD}}{F}\right)^2 e^{-7/F^2} e^{-\text{OD}/F}.
\label{efficiency}
\end{align}
OD is the optical depth of the medium, which we measure to be OD=1.3 in the waveguide used here.

We store weak coherent pulses (mean photon number of $\bar{n} = 1 $) in AFCs prepared with different storage times. We use pulses with the same temporal extent (and thus the same bandwidth) of the burn pulses and wait 15~ms after the comb preparation to escape spontaneous emission noise from the optical decay over the 2.85~ms Er lifetime. The first detected pulse corresponds to the unabsorbed transmission, which is not delayed, and then the echo is emitted at the designated time later. Figure \ref{fig:QM}c shows echoes for storage times from 300~ns to 1.2~$\mu\rm{s}$.  Calibrating the echo counts to the input photon number yields efficiencies of $\eta = 2.9\%$ for a delay of 300~ns and $\eta = 0.7\%$ for 1~$\mu$s and a noise figure (defined as noise added per photon retrieved) below 0.2 photons at all storage times shown. We note that the reported efficiency values correspond to the internal storage efficiency and do not account for input–output coupling losses in the device. The reduction in efficiency at longer delays arises from the decrease in the comb finesse at smaller tooth spacing. The enhanced efficiency at 300~ns may result from the comb spacing being commensurate with the
erbium superhyperfine structure \cite{thiel2010optical, sharma2023photon} (Figure S2). Consequently,  changing the superhyperfine spacing by changing the magnetic field should enable tuning which storage time sees this boost in efficiency. The measured efficiency for 300~ns storage is approaching the theoretical limit of 5.1\% for our OD of 1.3 and our finesse of 2.5. Increasing the efficiency in a waveguide configuration can be achieved by increasing both the OD and the comb finesse, with the optimal condition occurring at $\text{OD}=2F$ for a given finesse (Eq. \ref{efficiency}). The OD can be increased by more than a factor of 10 without substantial added propagation loss ($<$1 dB/cm) by using longer waveguides. Increasing the finesse, especially in much longer waveguides, would likely require a combination of improving laser stabilization, optimizing hole burning sequences, and finding a way to burn that is not via the waveguide itself (i.e. via a free space field incident on the chip from above). Furthermore, utilizing isotopically purified \textsuperscript{167}Er could enable the use of hyperfine states for more efficient burn sequencing. A finesse of 4 has been demonstrated in isotopically purified \textsuperscript{167}Er:YSO~\cite{PhysRevResearch.3.L032054}, which, if achievable in Er:TFLN, could result in an efficiency of 20\% or greater at the optimal optical depth as described earlier. Achieving even higher efficiencies requires coupling to an impedance matched cavity, where AFC storage with 80\% efficiency has been demonstrated in a europium-doped material \cite{meng2026efficient}. 

\begin{figure}
\includegraphics[width=\linewidth,keepaspectratio]{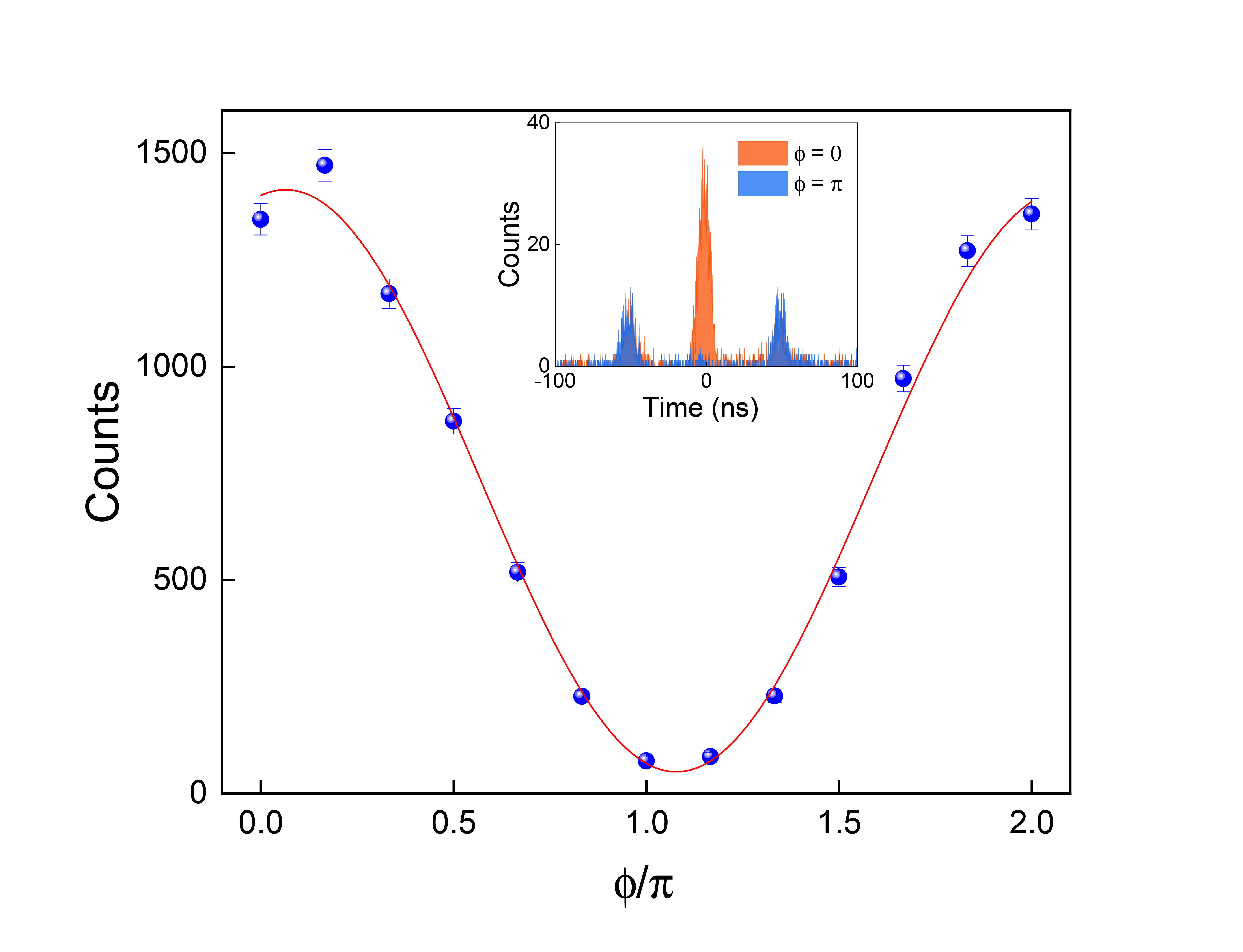}
\caption{\label{fig:Fidelity}  \textbf{Coherent Storage.} Retrieved photons in the central time bin as a function of phase between the early and late components of the stored time-bin encoded pulse with mean input photon number $\bar{n}=0.9$. Error bars represent Poissonian statistical uncertainties. Visibility is extracted from the fitted sinusoidal function (red line). Inset shows the output pulses at the points of maximum constructive (orange) and destructive (blue) interference.}
\end{figure}

For operation as a quantum memory, it is essential that the interface preserves the phase of the incoming photons in addition to not adding noise at the single photon level. To assess the coherence of the storage process, we store time-bin qubits encoded in weak coherent pulses ($\bar{n}=0.9$). The early and late pulses are separated by 50~ns and the qubit is analyzed by implementing partial readouts at different times \cite{de2008solid, staudt2006fidelity}. This is achieved by preparing two superimposed atomic frequency combs with storage times of 300 and 350 ns. The burn powers of the two AFCs were calibrated so that the two echoes had nearly the same efficiency. The resulting echoes are emitted at three distinct times separated by 50~ns, exhibiting interference between the early and late input modes in the central retrieved time bin. We observe high visibility interference fringes when we vary the relative phase, $\phi$, between the early and late pulses (see Figure \ref{fig:Fidelity}). We extract a visibility of 96.6~$\pm$~1.9\% from the fitted interference, demonstrating the phase preservation of the storage process. This sets the lower bound on our fidelity ($\mathcal{F}$)  which well exceeds the classical limit of $\mathcal{F} >2/3$  \cite{craiciu2019nanophotonic}. The visibility is mainly limited by small variations in the storage efficiency between the two AFCs and residual fluorescence noise from the hole burning procedure. 

\subsection*{Multimode and broadband storage}
\begin{figure}
\includegraphics[width=\linewidth,keepaspectratio]{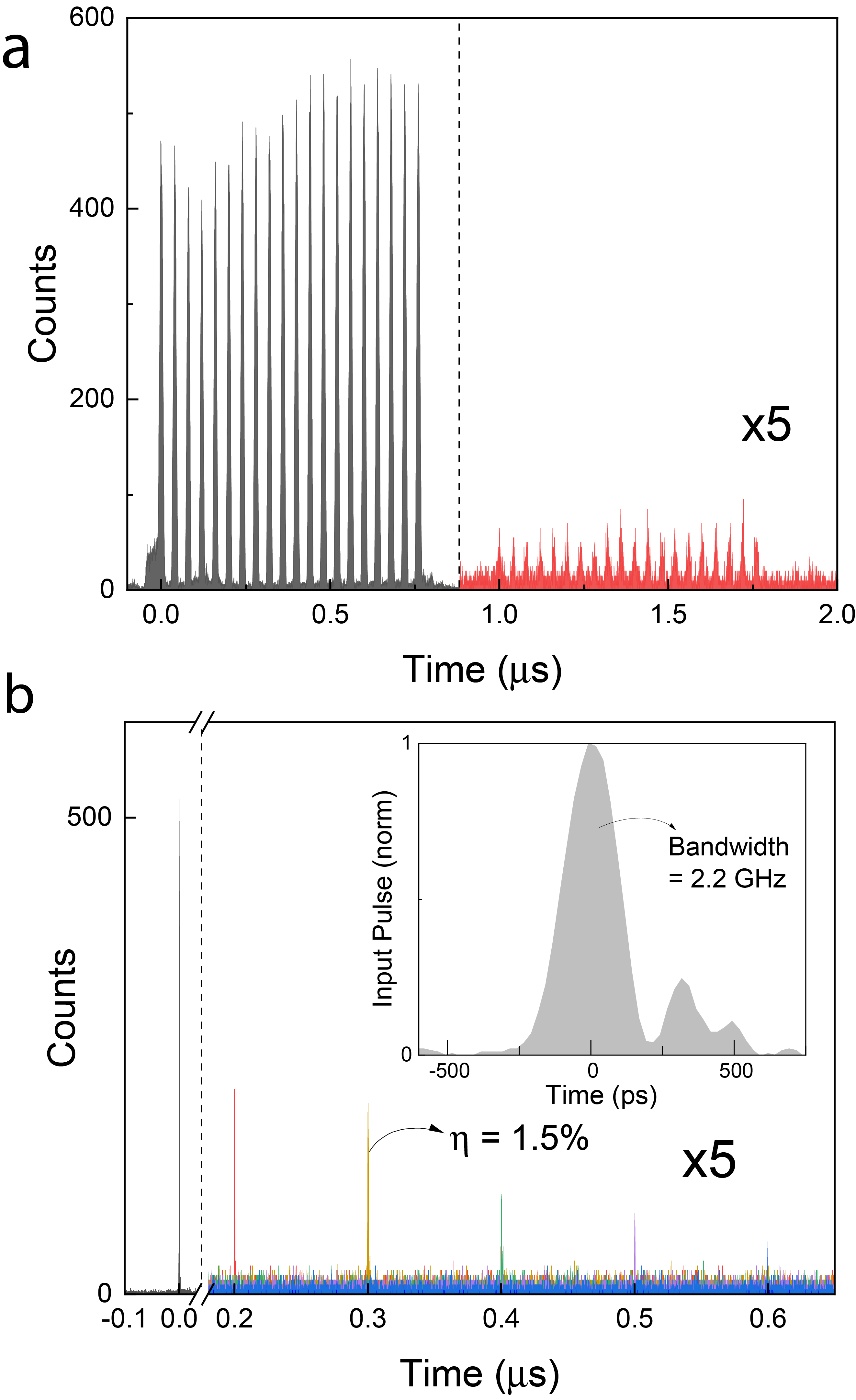}
\caption{\label{fig:Broadband}  \textbf{Multimode and Broadband Storage.} (a) Storage of 20 temporal modes for 1~$\mu$s. Stored pulses have a FWHM = 8.7~ns and $\bar{n}=1$, same as those in Figure \ref{fig:QM}c. (b) Storage of a 200-ps input pulse (inset shows temporal shape with $\bar{n}=1$) with a bandwidth of 2.2 GHz for different storage times. The undelayed leakage light is shown for reference at time zero. }
\end{figure}

An important advantage of the AFC protocol is its intrinsic multimode storage capability. To demonstrate this property, we store a train of 20 pulses separated by 40 ns in an AFC prepared with a storage time of 1~$\mu$s. Figure \ref{fig:Broadband}a shows that the corresponding AFC echoes are clearly resolved at the output with an average retrieval efficiency of 0.5\%. The maximum number of temporal modes that can be stored is given by the ratio of the storage time to the duration of an individual input pulse \cite{afzelius2009multimode}, 115 modes for this implementation, although in practice we were limited by our pulse generation electronics. 

Finally, we also showcase broadband AFC by storing a pulse with a bandwidth of 2.2~GHz. We prepare the broadband AFC via the previously described scheme, but with pulses with a FWHM of 200 ps  (see Figure \ref{fig:Broadband}b inset) generated using a home-built pulse generator to drive the intensity modulator. We store weak coherent pulses with $\bar{n} = 1$, achieving an efficiency of approximately 1.5\% at 300~ns with an added equivalent noise of  0.1 (see Figure \ref{fig:Broadband}b). The lower storage efficiency in the broadband regime is likely due to a combination of insufficient peak power for hole burning and increased leakage light deforming the comb. More laser power and a modulator with better extinction should limit this effect and enable similar efficiency to the 8.7~ns pulse storage. In principle, it should be possible to store nearly 3000 pulses in a multimode configuration, based on the ratio between the maximum storage time ($\sim$ 600 ns) and the pulse width. Furthermore, this 2.2~GHz bandwidth is currently limited by the speed of our pulse generator and the available optical power. Increasing both would likely enable broader band operation, with the ultimate limit set by a combination of the 160~GHz absorption bandwidth and the properties of the Zeeman split erbium ground state \cite{saglamyurek2016multiplexed}. 

\section*{Discussion}
We have demonstrated photonic quantum storage over one microsecond in a scalable nanophotonic platform. Our few-\% efficiencies are many orders of magnitude beyond what is possible with on-chip delays at this scale. The intrinsic telecom band operation is compatible with long-distance qubit teleportation, and the flexibility of the thin-film lithium niobate platform could enable the construction of multiple interconnected devices on a single chip, providing a viable pathway towards repeater based quantum networks. Additionally, a unique opportunity is the in situ integration of an entangled photon source with the AFC within the same TFLN platform \cite{zhao2020high}. This could provide a significant advantage for entanglement generation \cite{lau2025efficient}.

Apart from their role in quantum networks, on-chip optical delays offer significant advantages for quantum information processing. Efficient delay elements can enable time-delayed feedback, which reduces resource requirements for generating large-scale cluster states \cite{pichler2017universal, larsen2019deterministic}, and can also facilitate deterministic photonic quantum computing \cite{bartlett2021deterministic}. Our work presents a significant step toward implementing low-loss on-chip delays for photonic qubits beyond a few nanoseconds.

A key advantage of our platform is the ability to tune the optical depth (OD) by increasing the waveguide length, thereby enabling higher storage efficiency. Nevertheless, this approach alone is fundamentally limited to a maximum efficiency of 54\%. An alternative strategy is to employ high-Q resonators impedance-matched to the prepared AFC, which could in principle enable near-unity efficiency \cite{afzelius2010impedance}. Quality factors exceeding $10^7$ have been showcased in TFLN \cite{zhu2024twenty}, opening the door to high ensemble cooperativity and correspondingly enhanced efficiency. Furthermore, integration with fast electro-optic switching and multiple delay segments could enable dynamically tunable, on-demand delay control.

Altogether, this work represents a major step toward scalable and practical photonic quantum information processing. The combination of a leading telecom-band quantum emitter, erbium, with a leading quantum nanophotonic platform, lithium niobate, opens a wide range of possibilities for integrating quantum memory with photonic devices, potentially enabling wafer-scale fabrication and integration.

\textit{Note added} - In preparing this manuscript, the authors were made aware of a related pre-print showing similar results. \cite{wang2025storagetelebandtimebinqubits}

\section*{Methods}
\subsection{Device design and fabrication}
We used 625 nm thin film lithium niobate (z-cut) with a 0.5\% naturally abundant erbium concentration (NanoLN) doped during Czochralski growth. The waveguide was fabricated by etching 325 nm with argon ICP-RIE, cleaned using RCA and annealed at 500$^{\circ}$C for two hours in an O$_2$ environment. The waveguide is about 600 $\mu$m long and 0.8 $\mu$m across (Figure 1b). TE polarized light was coupled in and out using grating couplers, and we used angle-cut fibers glued to the chip. We measured a 5\% coupling efficiency per facet. 

\subsection{Experimental Setup}
The device was characterized in a 30 mK dilution fridge (Leiden) with a single-axis vector magnet used to provide a 1.5 T magnetic field along the c-axis of LN (Supplementary Figure 1). We utilized a tunable telecom laser (Toptica) whose polarization was controlled using a fiber polarization controller. The light was split into two paths: The first is the burn path, with the intensity modulator (iXblue) allowing us create short pulses ($<$10 ns) and the second path was probe scan path with an AOM used for scanning and probing our comb. The rest of the AOMs were used for gating and scanning as needed. The light collected from the device passes through another gate AOM, before being detected by a superconducting nanowire single photon detector (SNSPD) (PhotonSpot).

\section*{acknowledgments}
We acknowledge helpful discussions and support from Edo Waks, Uday Saha, Sanjukta Kundu, Santiago Vargas-Daniels, Nathaniel Irwin, Jason Huang, and Chris Anderson. This work was supported by the Navy, Office of Naval Research (Grant No. N00014-25-1-2394). This material is based on work supported by the U.S. Department of Energy, Office of Science, National Quantum Information Science Research Centers under Award Number DE-FOA-0002253, with additional measurement support provided by the Argonne Quantum Foundry through Q-NEXT. We also acknowledge the use of QUIUC-Net links funded by the Materials Research Laboratory at the University of Illinois Urbana-Champaign and the NSF Quantum Leap Challenge Institute on Hybrid Quantum Architectures and Networks (Award No. 2016136).
\vspace{1em}
\section*{Author contributions} 
P.B. and D.C. contributed equally to this work. E.A.G., and P.B. proposed the experiment. P.B. and D.C. conducted the experiments and analyzed the data. D.C., L.H., and E.C. fabricated the devices.  P.B., D.C., A.P., H.K., J.A., V.N., J.Z., A.M.D., and P.W. assisted in constructing the setup. E.A.G supervised the project. All authors contributed to the writing. 

\section*{Competing Interests} 
The authors declare no competing interests.

\bibliography{apssamp}

\clearpage
\onecolumngrid
\appendix
\section*{Supplementary Information}
\renewcommand{\figurename}{Fig.}
\renewcommand{\thefigure}{S\arabic{figure}}
\renewcommand{\theequation}{S\arabic{equation}}
\setcounter{figure}{0}
\setcounter{equation}{0}

\renewcommand{\figurename}{Fig.}
\renewcommand{\thefigure}{S\arabic{figure}}
\renewcommand{\theequation}{S\arabic{equation}}

\subsection{Experimental Setup}

\begin{figure}[H]
\includegraphics[width=\linewidth,keepaspectratio]{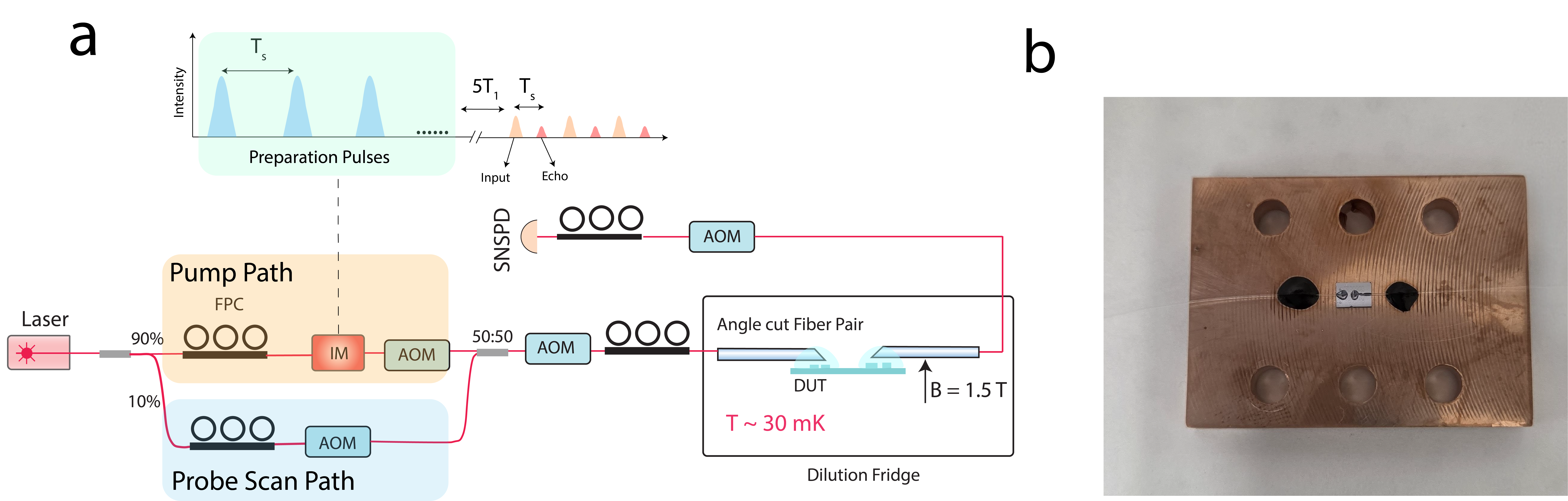}
\caption{\label{fig:Setup} (a) Schematic of the measurement setup. The acronyms denote AOM: acousto-optic modulator; IM: intensity modulator, with the top inset showing the AFC preparation pulse and input pulse sequence; FPC: fiber polarization controller; SNSPD: superconducting nanowire single-photon detector; and DUT: device under test.  (b) An angle-cut fiber is glued to the device using UV-curing epoxy, with Stycast applied to the sides to provide strain relief. }
\end{figure}

\subsection{Optical T$_{2}$ Measurements} 
\begin{figure}[H]
\includegraphics[width=\linewidth,keepaspectratio]{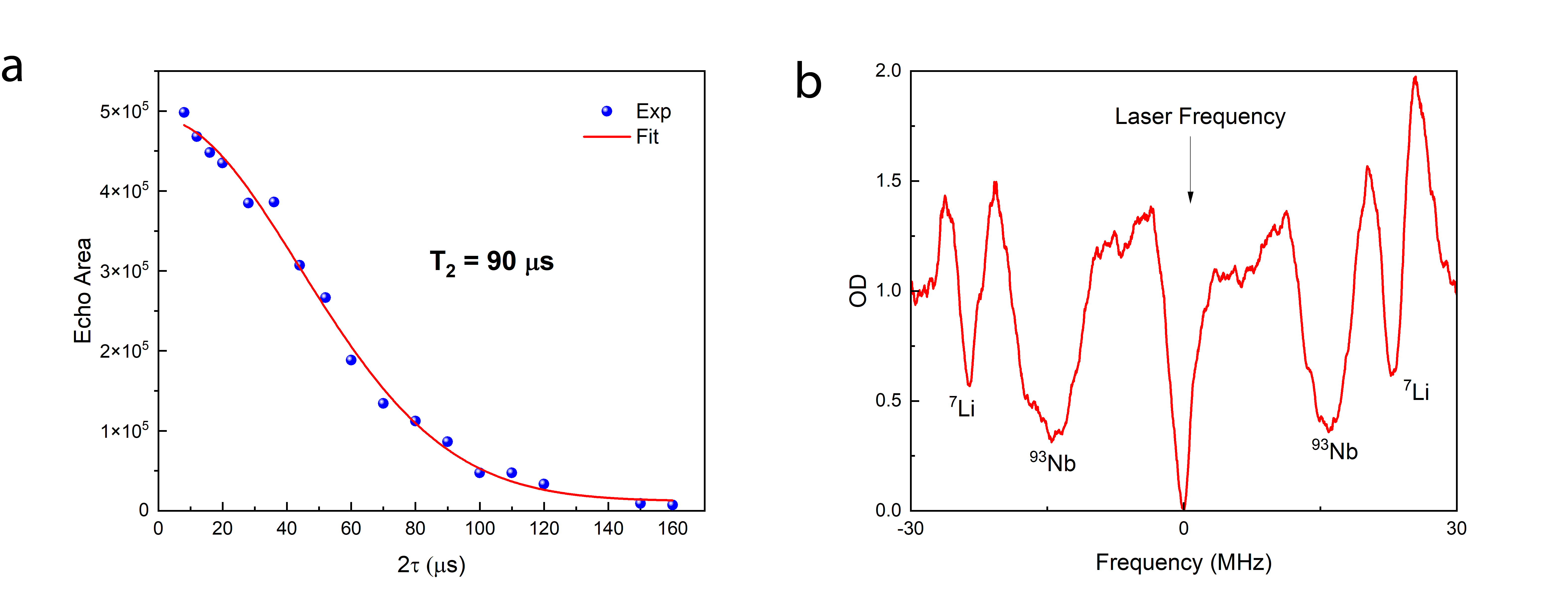}

\caption{\label{fig:PhotonEcho} The optical coherence time was determined using photon echo measurements (a) shows the photon echo intensity as a function of the pulse separation $\tau$ between the $\pi/2$ and $\pi$ pulses. (b) Transient spectral hole burning  showing the superhyperfine structure due to the coupling of the $\text{Er}^{3+}$ with the neighboring \textsuperscript{7}Li and \textsuperscript{93}Nb nuclear spins \cite{thiel2010optical} }
\end{figure}

Two-photon echo is a standard technique to extract the optical coherence time ($T_2$). Two short pulses of 100 ns and 200 ns were separated by a time delay $\tau$, which resulted in a coherent echo signal after time $\tau$ (Figure \ref{fig:PhotonEcho}). By fitting the echo peak intensity $I(\tau)$ to the expression \ref{eqn:Photon Echo} \cite{thiel2010optical},  

\begin{equation} 
\label{eqn:Photon Echo}
I(\tau) = I_0 \exp \left[ -2 \left( \frac{2\tau}{T_2} \right)^x \right]
\end{equation}
Where $x > 1$ is a measure of the spectral diffusion, we extract $T_2 = 90 \pm 4 \text{ $\mu$s}$.

\subsection{Optical Depth Measurement and Noise estimates} 

Our coupling efficiencies were calibrated by determining the transmission through the waveguide without erbium absorption. This was accomplished by sending progressively higher powers through the waveguide and measuring the transmitted power. The maximum transmission was measured to be around 0.26\%, corresponding to a near 5\% coupling efficiency per facet. This is the limit at which the erbium absorption has been saturated to transparency, leaving only the bare device transmission. A variable optical attenuator (VOA) was then set to this total transmission, and the same pulse sequence was sent through the device to give the amount of light transmitted in the absence of erbium absorption.

The average input photon number per pulse, $\bar{n}$, was determined by dividing the total number of detected counts per cycle by the number of pulses sent per experimental run. This yields the average photon number per pulse arriving at the SNSPD. To infer the average photon number coupled into the waveguide, this value was corrected for the single-facet coupling efficiency (5\%), the gate AOM transmission (40\%), and the SNSPD detection efficiency (60\%). We then varied the RF drive power of the AOM to calibrate the $\bar{n}=1$ operating point. The total added photon noise, $n$, was estimated by measuring the total detected counts per run within the temporal window corresponding to the output pulse, in the absence of the input probe. Since the effective added noise must be referenced to the input of the device, we estimate the noise added per retrieved photon as $n/\eta$, where $\eta$ is the AFC storage efficiency.

The optical depth was estimated by sending the same pulse sequence through both the device and the calibrated VOA, giving us an OD estimate of 1.31(1). We further validated this via spectral hole burning by measuring the ratio of transmission at the center of the hole to those outside it (Supplementary Figure S2b).

\section*{References}
\begin{enumerate}
  \item C.~W. Thiel, R.~M. Macfarlane, T.~B{\"o}ttger, Y. Sun, R.~L. Cone, and W.~R. Babbitt, \textit{Optical decoherence and persistent spectral hole burning in Er$^{3+}$:LiNbO$_3$}, Journal of Luminescence 130(9), 1603--1609 (2010).
\end{enumerate}

\end{document}